# Challenges in cybersecurity: Lessons from biological defense systems

Edward Schrom[a], Ann Kinzig[b,c], Stephanie Forrest[d,e,f], Andrea L. Graham[a], Simon A. Levin[a*], Carl T. Bergstrom[g], Carlos Castillo-Chavez[h], James P. Collins[b], Rob J. de Boer[i], Adam Doupé[e,j], Roya Ensafi[k], Stuart Feldman[l], Bryan T. Grenfell[a,m], J. Alex Halderman[k,n], Silvie Huijben[b], Carlo Maley[p,d], Melanie Moses[r,s,f], Alan S. Perelson[t,f], Charles Perrings[b], Joshua Plotkin[u], Jennifer Rexford[v], Mohit Tiwari[w]

[a] Department of Ecology and Evolutionary Biology, Princeton University, Princeton, NJ 08544, [b]School of Life Sciences, Arizona State University, Tempe, AZ 85287, [c]Julie Ann Wrigley Global Institute of Sustainability, Arizona State University, Tempe, AZ 85287, [d]Biodesign Center for Biocomputation, Security and Society, Arizona State University, Tempe, AZ 85287, [e]School of Computing, Informatics and Decision Sciences Engineering, Arizona State University, Tempe, AZ 85287, [f]Santa Fe Institute, Santa Fe, NM 87501, [g]Department of Biology, University of Washington, Seattle, WA 98195, [h]Retired Professor, Arizona State University, Tempe, AZ 85287, [i]Theoretical Biology and Bioinformatics, Utrecht University, 3584 CH Utrecht, The Netherlands, [j]Center for Cybersecurity and Digital Forensics, Global Security Initiative, Arizona State University, Tempe, AZ 85287, [k]Department of Electrical Engineering and Computer Science, Computer Science and Engineering Division, University of Michigan, Ann Arbor, MI 48109, [l]Schmidt Futures, New York, NY 10011, [m]Princeton School of Public and International Affairs, Princeton University, Princeton, NJ 08544, [n]Center for Computer Security and Society, University of Michigan, Ann Arbor, MI 48109, [o]Center for Evolution and Medicine, School of Life Sciences, Arizona State University, Tempe, AZ 85287, [p]Arizona Cancer Evolution Center, Arizona State University, Tempe, AZ 85287, [q]Department of Computer Science, University of New Mexico, Albuquerque, NM 87131, [r]Department of Computer Science, University of New Mexico, Albuquerque, NM 87131, [s]Department of Biology, University of New Mexico, Albuquerque, NM 87131, [t]Theoretical Biology and Biophysics Group, Los Alamos National Laboratory, Los Alamos, NM 87545, [u]Department of Biology, University of Pennsylvania, Philadelphia, PA 19104, [v]Department of Computer Science, Princeton University, Princeton, NJ 08540, [w]Department of Electrical and Computer Engineering, University of Texas, Austin, TX 78712

**\*Corresponding Author:** Simon A. Levin
Department of Ecology and Evolutionary Biology, 106A Guyot Hall, Princeton University, Princeton, NJ 08544, USA, (609) 268-6880, slevin@princeton.edu

**Simon A. Levin**
ORCID No.: 0000-0002-8216-5639
**Carl T. Bergstrom**
ORCID No.: 0000-0002-2070-385X
**Carlos Castillo-Chavez**
ORCID No.: 0000-0002-1046-3901
**James P. Collins**
ORCID No.: 0000-0003-1022-9952
**Rob J. De Boer**
ORCID No.: 0000-0002-2130-691X
**Bryan T. Grenfell**
ORCID No.: 0000-0003-3227-5909
**Stephanie Forrest**
ORCID No.: 0000-0002-5904-1646
**Andrea L. Graham**
ORCID No.: 0000-0002-6580-2755
**Silvie Huijben**
ORCID No.: 0000-0002-3537-1915
**Ann Kinzig**
ORCID No.: 0000-0001-7265-0289

**Carlo Maley**
ORCID No.: 0000-0002-0745-7076
**Melanie Moses**
ORCID No.: 0000-0002-8848-9554
**Alan S. Perelson**
ORCID No.: 0000-0002-2455-0002
**Charles Perrings**
ORCID No.: 0000-0002-4580-3697
**Joshua Plotkin**
ORCID No.: 0000-0001-6232-7613
**Edward Schrom**
ORCID No.: 0000-0002-1793-6433

**Classification**
Computer sciences, evolution

**Keywords**
Natural defenses, cybersecurity, immune systems, complex adaptive systems

**Abstract**

We explore the commonalities between methods for assuring the security of computer systems (cybersecurity) and the mechanisms that have evolved through natural selection to protect vertebrates against pathogens, and how insights derived from studying the evolution of natural defenses can inform the design of more effective cybersecurity systems. More generally, security challenges are crucial for the maintenance of a wide range of complex adaptive systems, including financial systems, and again lessons learned from the study of the evolution of natural defenses can provide guidance for the protection of such systems.

**Introduction**

The challenges faced in protecting biological organisms from natural enemies, and computer systems from malicious threats have much in common. In both systems, it is predictable that there will be security challenges, but unpredictable exactly what they will be, and when they will occur. Thus, comparing evolved and designed defense systems in these two arenas can inform our understanding of both, and perhaps lead to new ideas for improving cybersecurity systems. Security challenges are characteristic of many complex adaptive systems (CASs), in which patterns at high levels of organization emerge from localized interactions and selection processes operating on diverse agents at lower levels of organization, and feed back to affect those lower-level processes (1).

Through natural selection, biological systems have evolved mechanisms to deal with such threats, and it is instructive to ask what lessons can be learned for addressing cybersecurity issues. Focusing on both individual and population-level defenses, our aim in this paper is to identify commonalities, differences, gaps, and outstanding questions across these systems— naturally occurring defenses and those engineered to protect human populations and computer systems. This will require consideration of evolved, self-organized and designed defense mechanisms.

Securing cyber-systems is one of the central challenges of the 21st century. Failure to do so could lead to the demise of democratic governments, inadvertent nuclear war, or a severe disruption of financial markets, among many potential catastrophes. Over the past five decades, cybersecurity experts have made tremendous progress enhancing the security of computers, other digital devices, and networks, but the attackers always seem to be one step ahead with innovative methods for exploiting the latest defense.

Of course, by comparison natural biological defense mechanisms are remarkably diverse and sophisticated, since organisms have evolved in the presence of adversaries since their inception. These include the production of toxins, spines, antibiotics, sexual reproduction, and other mechanisms to deal with predators, herbivores, competitors and other agents, although our focus in this paper is largely on parasitic threats. Considering only individual protection of vertebrates, immune systems have evolved over several hundred million years in a wide range of host species, to protect against "attackers" (pathogens) that can range from viruses less than 10nm long to tapeworms exceeding 10m (2).

At the population level, societal public health strategies emerged early, although it is difficult to pinpoint precisely when. Many ancient religious injunctions helped prevent the spread of infectious disease in populations, suggesting that humans have been designing, adapting, and using population-level protections for a few thousand years.

National Medalist of Technology Carver Mead is often credited with saying that "As engineers, we would be foolish to ignore the lessons of a billion years of evolution." Cybersecurity and biological security must focus both on individuals (a single human, a single computer[1]) and populations (groups of humans, networks of digital devices). Both must guard against attacks by diverse, dynamic, and perpetually novel attackers. Both must learn to recognize what is "foreign" against a background of a potentially intricate "self" (many different kinds of cells, many different kinds of codes and use cases). Both must guard against attacks spreading through the population, without the defense being so draconian that it kills or incapacitates the host or catastrophically undermines the functioning of the system, such as might happen with autoimmune disorders, or 'protected' digital devices unable to execute legitimate code.

Looking to biology, particularly immunology, for insights about computing and security is not a new idea, and Wooley and Lin (3) provide a comprehensive survey. As one example of this earlier work, we consider intrusion detection systems (IDS), which are systems that monitor a computer or network for malicious activity or policy violations. This could include everything from buffer overflow attacks to hijacks of running programs to distributed denial-of-service attacks to the recent SolarWinds attacks. Prior to 1996, IDS used the same principle as virus scanners, known as signature detection. Signature detectors consist of a database of known intrusive patterns—the signatures—which are actively compared to system activity, for example, to detect port scans and other forms of network attacks (4, 5). In 1996, the first practical anomaly-detection system was introduced, patterned explicitly on the immune system (4, 5). This work changed the paradigm of IDS from one that recognized pre-programmed patterns to one that automatically learned the normal behavior of the system (self) simply by observing its operation and responding adaptively to unfamiliar patterns. Concretely, the system learned patterns of normally occurring short sequences (n-grams) of system calls (the mechanism by which executing programs access resources such as writing to a screen or opening a file). False positives are a concern in IDS, for both signature (6, 7) and anomaly detectors, just as autoimmunity is for immune systems. Following the immune system's use of second signals, e.g., costimulatory receptors on T cells, a similar mechanism was added to the system-call approach in a system called pH (process Homeostasis) (8). In subsequent anomaly-detection systems, which focused on network IDS, even more immune-like mechanisms were added, mimicking negative selection of detectors (for flexibility and distributed execution), a secondary response (so that previously seen attacks are responded to more quickly and aggressively), diversity of MHC presentation (to avoid single points of failure), and avidity (to control false positives) (9).

Recent advances in biology, e.g., the discovery of CRISPR/Cas and its role in defense, together with rapid changes in the computing ecosystem (e.g., high-speed networking; widespread adoption of cloud computing; business standardization on a few enterprise platforms; the roll-out of Internet-enabled devices, many with significant autonomy; machine learning; social networking; and the rapid acceleration of cybercrime and nation-state cyber conflict), suggest the value of a fresh look at how lessons from biology might be leveraged to better protect these systems.

Of course, there are important distinctions and differences as well. Nature has created defense 'solutions' through the process of evolution by natural selection. In contrast, today's cybersecurity systems are engineered iteratively as new threats arise, leading to systems designed with intent but not necessarily intelligent design. Public health is somewhere in between, a mixture of evolution and engineering: Pathogens and immune systems evolve and adapt, but defensive measures of outbreak control are engineered at the system level, as we are witnessing in real-time with the COVID-19 pandemic. Natural selection operates primarily at the level of the individual, with weaker population-level selection, and consequently immune systems primarily respond to threats against individual organisms. Public health is primarily concerned with population-level protection. The targets of engineered cybersecurity systems, in contrast, could be either individual devices or networked populations. Natural selection is limited by mutation and recombination rates and the incremental nature of genetic change across generations, while cyber-engineers and public health designers can, in principle, make wholesale changes and longer leaps, potentially exploring a larger 'solution space' (though realized changes, like in biology, are often subject to historical constraints). Finally, to some extent, natural selection favors solutions which deal with problems that have been experienced, even if the solutions are bound to become burdensome in the future (10), whereas the rational design of cybersecurity or public health has, in principle, the benefit of foresight (see Box 1), although many measures that have been implemented suffer the same burdens as those that saddle natural selection processes.

---

[1] We will often use 'computer' as a shorthand for computing devices, including routers, servers, smartphones, etc.

In this paper, motivated by these considerations, we address two broad questions:

> 1. Do organismal and cyber defense systems suggest general principles of optimal defense strategies?
> 2. Can insights from immunology suggest improved cybersecurity, and can insights in cybersecurity inform public health?

In the subsequent sections, we provide a framework for studying the problem of defense across cyber and living systems, drawing meaningful connections between them and abstracting general principles where possible. Although we focus primarily on biological and cyber defense, we discuss public health examples when they are particularly relevant.

The paper concludes with a set of cross-cutting future research questions, which are intended to transcend mere comparisons of systems and inspire general principles. We provide the analysis in three general areas:

> 1. Framing the context of defense,
> 2. Choosing defense strategies, and
> 3. Evaluating performance.

## I. Framing the Context of Defense

The defensive context includes the goals of defense, the goals of attack, and the environment in which attacks and defense occur. All three of these components should be considered when searching for common lessons among different defensive systems. Below are specific considerations for framing the context of defense, along with examples from cybersecurity, biological immunity, and public health.

**The Goals of Defense.** In biological immunity, natural selection produces defensive systems that enhance the lifetime reproductive output of the organism being defended. There is no selection for host comfort, or post-reproductive survival. This helps explain why the mammalian immune system senesces (e.g., experiences inflammatory malfunctions, and/or permits cancer) in old age, after reproduction is complete (e.g., 11). Selection for reproductive output rather than health may also explain why some parasites are eradicated from the body, whereas others are allowed to persist as other mechanisms mitigate their negative impacts (e.g., evolved tolerance: 12, 13).

Cybersecurity defense, in contrast, is not focused only on successful 'reproduction' of the defended host and nearly always has to deliver on a broader array of goals. For example, in addition to repelling attacks, effective computing systems must also maintain a comfortable user experience while safeguarding data and results for a reasonable cost. Strategic defense strategies may also entail 'sacrificing' initially infected[2] individuals, e.g., by disconnecting them from a network, killing processes or revoking access, to protect populations of computing devices. (In some cases, an infected system may be permitted to continue while being prevented from doing harm by network limitations or sandboxing.) Public health defense has similarly diverse (and potentially conflicting) goals, in terms of the protection of individuals versus the protection of the many. (As an extreme case, escapees from quarantined ships were routinely executed.) Indeed, in non-human systems, for example, in the face of foot-and mouth disease, infected individuals as well as non-infected ones on the same or adjacent farms may be culled to protect the population, and quarantining measures in human populations may more subtly introduce similar tradeoffs, for example, the original quarantine facilities (lazarettos) were culling mechanisms in practice.

**The Goals of Attack.** As with defense, evolution by natural selection in biological systems selects for the reproductive capacity of 'attackers.' This can either be reproduction (proliferation of pathogen) within a host, or reproduction based on transmission between hosts. If a pathogen kills its host 'too quickly,' between-host transmission is impeded, leading in many cases to the evolution of intermediate virulence (14, 15). In some cases, such as Ebola, transmission can occur after death, thus subverting selection for intermediate virulence. Furthermore, in other cases, like rabies, spread relies upon severe symptoms of infection, leading to a similar outcome (16). Pathogens can exploit (and induce) the defensive physiology of the host to spread—through, for instance, coughing (Influenza A) or vomiting (hepatitis). Pathogens have been known to manipulate host behavior in even more surprising ways to facilitate transmission. "Zombie ants," for instance, are ants infected with a fungus that induces the ant to climb up a plant and attach. The fungus then 'eats' the ant from within before producing spores that can be scattered from the elevated position (17).

Cyber-attacks may select for reproductive capacity, but generally it is in service of another goal, such as incapacitating a single host as in a targeted attack, incapacitating multiple hosts, stealing data or credentials (in which case incapacitating the host could be counterproductive, much as in intermediate virulence), undermining faith in a corporation or government, holding systems to ransom or subtly polluting them, spreading misinformation, or simply making mischief. Similar to viruses, cyber systems can manipulate 'host' behavior—since attackers can exploit both the behavior of the computer (causing it to execute foreign code), and similar to ants they can manipulate the human user of the computer (as in phishing attacks).

Transmission between hosts occurs when computers are tricked into transmitting damaging code or sending emails to all contacts in the host computer. People can be manipulated into clicking links that load malware. In amplification attacks, an adversary sends small requests to many benign hosts that trick each of them into sending large responses to overwhelm a target victim with excessive traffic.

**System Scale.** By definition, a CAS spans a range of temporal, spatial, and/or organizational scales, as do cyber systems, which can range in scale from a single transistor to billions of transistors on a single chip to large enterprise systems to the billions of computers and 'things' connected to the Internet.  In a CAS, interactions that cross these scales often contribute to the non-linearities responsible for surprising system behavior, and we expect similar phenomena in the cyber domain.  In particular, unanticipated effects can occur in defensive systems when the scales of the attack differ from the scales of defense. For example, localized molecular signals that induce inflammation are triggered by, and help kill, nearby bacteria, but if the same signals are spread throughout the body, they can cause septic shock and ultimately host death (18).

---

[2] Note that throughout we often speak of 'infection,' though not all cyber-attacks are the result of the spread of malware. Here, we use Infection in the cyber context to mean any attack that compromises all or part of a cyber system.

While mismatched scales are often a challenge to effective defense, they can also present opportunities for new defensive strategies.  At the population scale, heterogeneity among individuals may preclude a single widely effective attack strategy (see Box 2).  For example, heterogeneity among people in which MHC genes they possess, and therefore in which antigens they recognize, may prevent parasites from becoming antigenically invisible to (hidden from) any large fraction of a population.  Such heterogeneity may be maintained by negative frequency-dependent selection, whereby rare MHC alleles are favored because pathogens are not adapted to them (19).  Similarly, heterogeneity in operating systems may protect larger systems of networked computers (20), and several other forms of diversity have been proposed for leveraging diversity to improve security. Unfortunately, there has been considerable narrowing of choices (only a few operating system and chip designs account for almost all examples in the world.).  Probably the most widely deployed form of engineered diversity for computing is address-space randomization (8), which is shipped by default on most of today's programs and operating systems.  Similarly, in some wireless networks, frequency hopping—transmitting radio signals over rapidly changing frequencies known only to the legitimate parties to the communication—is an effective way to resist an adversary deliberately jamming the network.  More generally, a potential victim can confuse an adversary by changing many different aspects of the underlying system (e.g., addresses, names, software stacks) to dynamically shift the attack surface, confounding an adversary's efforts to plan attacks in advance (21).

Unanticipated effects can occur in defensive systems when defense strategies operating at different scales interfere with one another (10).  As an example, the implementation of two-factor authentication at the institutional level may actually compromise security at the level of the individual computer, because individual users may be lulled into being less vigilant about creating strong passwords (22).  In public health, a successful national vaccine campaign may lull individuals into skipping booster shots and engaging in riskier behaviors, resulting in local patches of susceptible individuals and a renewed risk of outbreak; this may be a factor, among many, including the recent resurgence of measles due to vaccine rejection based on a lack of perceived need.  Both public-health and cybersecurity systems face challenges of local (person-to-person, computers linked to a common server) and long-distance transmission.  Public-health interventions originated at a time when rapid long-distance transmission (e.g., via airplane travel) was less common, and those legacies may influence strategies today.  Cybersecurity, in contrast, has primarily evolved in the context of rapid long-distance transmission.  The interventions required when one expects primarily local transmission are different from those when one expects long-distance transmission.  Public health containment strategies may need to evolve for an era of widespread long-distance travel, in which case they may converge to similar solutions as those that prove effective for large computer networks.  The time scales in the two systems are also different—biological viruses require at least several hours to travel halfway around the world while computer malware transmission can be nearly instantaneous. Progress could be made by categorizing the spatial and temporal scales of 'infection' in cyber-security and biological-health systems, and comparing best practices under common conditions.

**Third Parties in Cybersecurity and Organismal Defense.** Attackers and defenders are not always the only relevant actors in biological or cybersecurity systems. Intermediaries often dubbed "third parties" or "third actors" may wittingly or unwittingly benefit the attackers and/or the attacked. For example, in the 2016 elections, those who posted minority political opinions on social media became third actors when Russian hackers amplified their posts to distort public perception of the political climate. In the biological realm, pigs and birds are third actors who provide the habitats in which different strains of influenza can mix to create new strains that can cause pandemics in human populations.

As a first approximation, we can think of security threats as comprising three parties: those who threaten (party 1), those who are threatened (party 2), and bystanders—those who play an active role in enabling, exacerbating, or mitigating the threat (third parties). In both cybersecurity and organismal security, the harm to those who are threatened depends on a wider 'ecosystem' of actors. In both cases, risks are mediated by a hierarchical set of networks—the Internet and the market for goods and services are examples—secured at different points by different organizations. Some countries close the network to potentially politically risky inputs as well as many ordinary information leaks. Cybersecurity depends more on private organizations operating at the supra-national level, although that is changing as governments take a more active role. Public health depends more on public institutions operating largely at the national level. Each seeks to contain risks to the networks or sub networks in their charge. Third parties whose actions affect risk include a wide range of private interests: e.g., software developers, device providers, those who trade in risky materials, and those who travel along risky routes. By extension, third parties also include the species that mediate vector-borne or zoonotic diseases or that compromise the management of epizootic diseases.

*The Cybersecurity Ecosystem.* Those who threaten harm in cybersecurity systems, "attackers," include the organizations and individuals who plan attacks, as well as those who execute or launch them. Ancillary actions include:

> 1. probing for weak spots (e.g., unprotected network ports, easy passwords);
> 2. analyzing code for failure modes (e.g., that induce stack overflows);
> 3. curating exploits (maintaining a library of vulnerabilities or code exploits, running a marketplace or a sales operation);
> 4. sharing information (dark web advice sites and information sharing venues);
> 5. operating testbeds to verify exploits, or honing the use vectors;
> 6. subverting resource managers (money laundering, financial access, loans);
> 7. recruiting of vectors (network access with high fire power, creating botnets of compromised machines for massive or secretive attacks, etc.);
> 8. planting hardware weaknesses (trapdoors, advanced silicon attacks) when hardware is designed or manufactured; and
> 9. planting software weaknesses (compromised programmers introducing subtle bugs intentionally).

Those who are threatened comprise the people benefitting from the system being attacked, where the systems might include connected machines, insertable devices and media, and network wireless connections, and self-protection measures such as security devices (e.g., firewalls, trusted hardware enclaves, encryption engines); security software (e.g., anti-virus);

security systems and operations (intrusion detection, red teams, security specialists).

Third parties include tech providers (operating system vendors, storage and application servers, cloud providers); device providers (IoT, drone, sensors, appliances, autos and the hardware/software/operations they provide); network providers (ISPs, classic communications firms, wireless); curators of risks (e.g., blacklists from government supported sources such as CERT or from private companies such as security firms or major technology companies); and financial providers (cyber insurance).

A recent example of a highly sophisticated supply-chain attack against this cyber ecosystem is the recent SolarWinds software attack, attributed to Russian operatives and detected by FireEye, a computer security company.  SolarWinds involved several carefully coordinated stages, some of which have analogs in the biological world.  This is, however, a strategically planned long-term operation, which would be difficult if not impossible to engineer through a series of mutations.  In brief:

> 1. Intruders obtained access to the computing systems at the company.  SolarWinds, probably using common industry code-sharing software.
> `
> 2. They infiltrated the mechanism used to build copies of the Orion network management software product.  Modern software is assembled using a 'build' process, roughly analogous to developmental processes in biological systems:
>> a. When a new software update was being produced, the intruder temporarily replaced key code, compiled it into the product, then replaced the usual versions so there was no apparent change.  This is a form of deception and planned self-repair that has no obvious biological analogue.
>> b. The compilation and assembly process was managed carefully and automatically monitored for errors or other flags that would signal the presence of the intrusion to the system or its operators.
> 3. Companies that purchased Orion received software updates regularly and trusted their validity.  But when they installed these special versions, they installed additional hidden and undesired capabilities.
>> a. The bugged system waited two weeks before taking actions.
>> b. It then communicated with outside systems for instructions.  It could export secret data, install other processes, or destroy the system or render it nonfunctional.
>> c. This was accomplished with a modified version of a common tool (Cobalt Strike) that security experts use to test their own systems (known as red teaming).

Parts of the SolarWinds attack are analogous to biological mechanisms.

2a. (deception and self-repair) is common among parasites, and most organisms have self-repair mechanisms, for example, the famous Cas9 system in bacteria.

2b. (monitoring for errors) is similar to the functions of the adaptive immune system.

3a. (dormant period) is common with infections—they can take a significant time to reach a level that causes damage or elicits a major response from the immune system.

3b. (external communication or internal damage) is similar to what occurs when a virus overwhelms a cell.

3c. the Teardrop malware manager was a modified tool used by the security organization, so it appeared to be "self," not an outside attacker, similar to the use of immunosuppressant chemicals or the way tuberculosis hides within macrophages.

However, the totality of the attack, with long malice aforethought and many indirect stages, seems more sophisticated than naturally occurring infections that we know of.

***The Organismal Defense Ecosystem***.  Infectious diseases of plants and animals, like infectious diseases of humans, are frequently transmitted by trade and travel, by exposure to infected food, wind, or water borne pathogens, or by disease vectors such as mosquitoes, ticks, fleas, flies, and sandflies.  All those whose behavior increases the risk of transmission of harmful infectious diseases bear some responsibility for causing harm, even if the activity that affects risk is legal.  The level of threat is frequently specific to the pathogen involved (its virulence), but can be increased by the failure to take preventive or precautionary actions (vaccines, personal hygiene, social distancing or separation, wearing masks) and the failure to implement biosecurity measures in the management of plants and animals.

Those who are threatened comprise the people directly or indirectly harmed by infectious diseases of humans, animals, and plants, and the defensive mechanisms available to them, including the immune system of directly harmed individuals together with any preventive or precautionary measures adopted.

Third parties are a mixture of positive, negative, and mixed influences.  They comprise the array of regulatory bodies, intergovernmental agencies (e.g., World Health Organization; World Trade Organization, World Animal Health Organization, and the Sanitary and Phytosanitary Agreement), non-governmental organizations (e.g. Gates Foundation, International Union for Conservation of Nature); national, regional, and local public health agencies (e.g., NIH, CDC, local vector control); national border inspectorates (e.g. APHIS); private interests engaged in activities that impact risk (in health, education, agriculture, land management, trade and commerce).  Third party interventions include both biosecurity measures that are risk reducing, but also omissions that are risk-increasing such as:

> 1. failure to disinfect trade or travel vessels (airplanes, ships, trains, buses, trucks, containers etc.);
> 2. failure to control disease vectors (vector habitats, vector abundance, vector competence);
> 3. failure to control disease contact zones (wildlife agriculture interactions, plant trade  entrepots); and
> 4. infection by traveling NGO, medical, and security personnel.

Box 3 provides case studies to illustrate the impacts of third parties (specifically vectors) in biological and cyber defense systems.

## II. Choosing Defensive Strategies

Immune systems and cyber defense systems offer a wide array of strategic options. Considering both the biological and cyber domains, we (and others; e.g., (23) on biological defenses) find that defensive strategies fall into five general categories:

> 1. Avoidance, or preventing contact with attacks.  Examples: shunning unsafe websites, quarantining sick individuals, social isolation.
> 2. Creating barriers, or preventing entry of an attack upon contact.  Examples: skin and mucus blocking pathogen entry, cryptography, firewalls to prevent malware entering a network, passwords and other access controls to prevent outsiders from using a computer.

3. Detection, or recognizing attack entry when it has occurred. Examples: antigen recognition by Toll-Like and T cell receptors, intrusion-detection system, virus scanners, honeypots.

4. Alleviation, or reducing the harm caused by an attack. Examples: switching off a compromised server, slowing down an anomalously behaving program, repairing damaged tissues or otherwise tolerating infection.

5. Counterattack, or eliminating the attack and/or its source. Examples: issuing a "takedown request" to remove counterfeit websites or launching a denial-of-service attack against suspected adversaries, and killing infected cells or shedding parasites.

A defensive system may deploy multiple strategies, as well as multiple versions or instances of the same strategy. Interactions among these layers determine overall effectiveness. The cyber principle of defense-in-depth says that adding defensive layers strengthens defense, and indeed, mammalian immunity implements multiple versions of all five strategies listed above. Simpler immune systems exist, however, which are also effective. For example, corals (24) and unicellular organisms such as bacteria (25) arguably achieve at least four of the categories of strategies. "Complexifying" defensive strategies comes with costs. Each new "layer" or strategy means existing mechanisms may be underutilized, and that may reduce selective pressure for continued adaptation. Strategies can also interfere with each other. For instance, barriers that must be breached (e.g., skin) may lengthen the time required for detection by other components of the defense system (e.g., phagocytic cells). Sophisticated multi-layer defense systems can arise by natural selection without providing any benefit over simpler defenses in the long run. Thus, a successful design strategy must carefully balance a number of factors when determining the number and arrangement of defensive layers. Below are specific considerations that affect the choice of defense strategy—including resource costs, cost of false positives, communication, and learning—along with examples from cybersecurity, biological immunity, and public health. We also include related future research questions.

**Resource Costs.** Maintaining and deploying any defensive system has costs. For example, antivirus software can increase run-time for legitimate software. Producing antibodies in response to infection uses proteins that could have been invested in growth or reproduction. Investing in public health uses tax dollars that could have gone to education. Ideally, the cost of a defensive strategy should be commensurate with the risks faced. The degree of risk, and therefore the incentive to implement a given defensive strategy, may vary across the agents in the system. For instance, the social impact of a public-health worker contracting an infectious fatal illness may or may not be greater than if residents of a nursing home were to be infected. Immune systems 'solve' the cost problem (imperfectly) through natural selection—the immune system investment in clearing or containing pathogens is predicted to be commensurate with the risks (26, 27). Immune systems, though, might 'underinvest' (fail to respond) in the face of novel pathogens, or 'overinvest' by unleashing an immune response that is more damaging than the pathogen (auto-immunity disorders; see below), suggesting that it is tricky to maintain an ideal balance. Public health and cyber defense systems must rely on forecasting (rather than natural selection) to determine which expenditures are warranted, and both systems can similarly suffer from over-or under-investment. Ironically, both types of defensive systems are often accused of overinvestment when their defenses are successful and disaster is avoided, e.g., companies often reduce their security investments after periods of relative calm. Similarly, it was only after the 2017 Equifax compromise, where information from 147 million people was leaked, that Equifax significantly invested money in cybersecurity.

Cyber- and public-health defenses have characteristics of an impure public good—agents benefit from their own defenses but also from the defenses of others. For example, vaccinating one person in an office against influenza or buying a single copy of anti-phishing software can reduce the risk of damage for his/her coworkers. This creates a 'free rider' problem, in which some agents wait for others to pay the costs of defense. This may lead to underinvestment in security measures, use of unencrypted communication networks, and anti-vaxxers.

**False Positives.** In both organisms and computers, some ingressions are dangerous, but the majority are not (e.g., food, e-mail messages, most software updates). Fighting innocuous ingressions can be as costly as permitting dangerous ingressions. As a result, detection strategies must be carefully tuned. False positives occur when the immune system attacks an innocuous substance or its own uninfected cells, or when a cybersecurity program denies authorized code the access it needs to continue executing. False positives in immunity (autoimmunity) are difficult to treat, often more challenging than treating infections, so we may wish to use medical interventions to decrease false positives in favor of treating false negatives. On the other hand, false positives in the cyber realm (e.g., legitimate emails withheld by a spam filter) are rarely tolerated by users, and most users favor a norm that tolerates some 'junk mail' in favor of reducing negatives.

Reducing false positives often increases false negatives (28). "Breaking" this tradeoff—so that one could simultaneously reduce false positives and false negatives—offers significant advantages in any defensive program. Potential approaches to solving this problem in both cybersecurity and public health could include: 1. cellular/machine learning, with individual computers or public-health bureaus updating their profile of expected attacks in real time; 2. majority voting, through which sensitivity can be maintained but false positives blocked; and 3. incubated tests, through which suspected attacks can be queried in a quarantining environment and permitted if they prove harmless. (However, attackers strive to circumvent incubators through cloaking, randomization, timing and external stimulus requirements, and other techniques.)

**Communication within Defensive Systems.** Deploying multiple defensive strategies, or duplicates of defensive agents in many places, can require communication among these components. The many thousands of nodes comprising modern computer networks continually exchange packets of information, and thousands of immune cells constantly use signaling molecules to alert other cells about the presence of danger. Every signaling pathway is an opportunity for subversion (29), such as spoofing or man-in-the-middle attacks. On the one hand, centralizing defensive agents and eliminating signaling pathways reduces the number of communication channels vulnerable to subversive attack. On the other hand, distributing defensive agents and adding signaling pathways reduces the impact of subversion when it occurs. It could be helpful to compare "signaling logics" that are hard to hack, both in biology and cryptography. In computing systems it is probably preferable to harden the communication pathway than the software that depends on it. In addition, some cyberattacks are much easier to detect based on the means of communication rather than the content of the message; for example, spam email may be more readily detected based on the IP address, geographic location, or time-of-day of the transmission rather than analyzing the message contents (30).

**Learning within Defensive Systems.** A fundamental problem in both immunity and cyber defense is that attackers have more frequent opportunities to update their strategies than do defenders. Cyber attackers can privately test many attack strategies before launching the best one, and parasites have much shorter generation times and larger effective population sizes than hosts. Defense systems composed of distributed autonomous agents can partially close this gap in evolution rate by allowing the agents to evolve as a single defensive response unfolds (e.g., positive selection for initially rare antigen-specific cells in mammalian immunity). When evolution within a defensive response is paired with memory of the attack, the defense system is capable of learning. With the growth of large, distributed enterprise systems, coupled with new technical trends towards very lightweight processes and numerous processors, it is easy to imagine a new kind of cybersecurity system, which leverages the advantage of large populations and on-line learning (9).

## III. Evaluating the Performance of Defense

However thoroughly designed or well-adapted a defense system may be, the unpredictability and continual evolution of new attacks means that monitoring and updating defense will always be necessary. In both cybersecurity and immunity, new defenses inspire new attacks and vice versa. Below are specific considerations for evaluating defense systems, along with examples from cybersecurity, biological immunity, and public health.

**Co-evolutionary Patterns and their Predictive Power.** The creation of new attacks in response to new defenses and vice versa is a coevolutionary process often called an arms race. In human-managed systems, the mere threat of an arms race—even without attack—can cause both attackers and defenders to escalate their strategies. Attackers may preemptively unleash a massive assault to overrun anticipated defenses, and defenders may invest in additional defense based only upon rumor of advancing attacker capability. Finely detailed data on each new attack strategy and defense counter may reveal instances of wasted resources (unnecessary escalation) or impending catastrophe (mild attacks giving way to a major attack). For example, time series of viral sequences and antibody repertoire sequences in HIV patients often reveal patterns that accurately predict how long the patient has to live. From publicly available data on security breaches and the Great Firewall of China to the entire field of Internet measurement, cybersecurity is teeming with similarly detailed data.

**Identifying Worst-Outcome Scenarios.** Coevolutionary patterns are likely to predict catastrophes only in a probabilistic sense. The magnitude of the worst-case scenario, even if its probability is quite small, is also relevant for the performance of a defense system. For example, in a population with a fully connected contact network, even if every individual is vaccinated, just one mutant strain can cause a devastating epidemic. Conversely, programming heterogeneity across devices comprising the Internet of Things could confer population-scale defense, but poor security and common design and origin in most of these individual devices may still render them susceptible to a large-scale attack with potential devastating ramifications for many people.

**Feedbacks on the Context of Defense.** If a defense system is carefully designed for optimality in its specific context, then unanticipated changes to that context may be the origin of catastrophe. Thus, vigilance in detecting changes to the context of defense is essential. Contextual changes can be externally driven. For example, shifting political alliances among nations and political groups may change the origin, and thus the resources available, for cyberattacks. Contextual changes can also be driven by the defense system itself, in the form of unintended consequences. If you lock your house will thieves try your neighbor (theory of constant threat) or reduce total crime? For example, using a gene drive system to cause local extinctions of malaria-carrying mosquitos may impose strong selective pressure for Plasmodium parasites that can survive in other vectors. Suppose another biting insect species with a much wider geographic range becomes a new vector. Now the context for defense has changed drastically—a new third actor is involved, the spatial scale of attack has changed, and many new populations of people are at risk.

**Future Research Directions and Opportunities**
This analysis raises far more questions than it settles. We list just a few here:

1. How do strategies to defend individuals and populations differ, and how can they conflict?
2. How might learning create new defenses but also vulnerabilities, and how can offense make strategic use of learning?
3. Can cyber and public health systems predict and preemptively act against prospective attacks?
4. Are tradeoffs between false positives and negatives fundamental in the biological or cyber worlds? If yes, are there principles for managing the tradeoff?
5. What are the necessary tradeoffs between complexity (multiple layers) and simplicity for both fallibility and cost? How does depth and breadth of communication channels affect vulnerability and defense?
6. Can defenses encourage less virulent or dangerous attacks, and not just by deflecting to other victims?
7. Can shifts in the ecosystem of defense, such as sudden inclusion of new third actors, be predicted?

**Conclusion: Moving Forward**
Whether a defense system has evolved or been designed, or is some combination of the two, we find that there are meaningful parallels in how the defensive contexts are framed, strategies chosen, and performances evaluated. Many such parallels have been spotted among cybersecurity, biological immune systems, and public health. Even more analogies between these two systems are certainly possible, but the most productive way forward is to use the existing analogies to suggest and study *general principles of defense against unpredictable attacks*. Lists of proposed principles already exist in both fields separately, with their generality across systems yet to be examined (e.g., 31). Moreover, we hope the open questions above will spark collaborative study, whether by sharing data and analytical techniques or constructing theoretical models, to identify more general principles. Finally, these general principles must be vetted in other realms, such as national defense against terrorism. Answering these open questions by development of general defense principles would constitute extraordinary advancements in our individual fields of study and in our cross-disciplinary understanding of complexity.

**Acknowledgements**
The authors would like to thank Arizona State University's College of Liberal Arts and Sciences for providing the funding for the workshops that led to this paper as well as Princeton University for hosting one of the workshops. Portions of A.S.P.'s work were done under the auspices of the U.S. Department of Energy under contract 89233218CNA000001 and were supported by NIH grants R01-AI028433 and R01-OD011095. S.F. gratefully acknowledges the partial support of NSF (CCF 1908633, IOS 2029696), DARPA (FA8750-19C-0003, N6600120C4020), AFRL (FA8750-19-1-0501), and the Santa Fe Institute. S.A.L. thanks the Army Research Office (W911NF-18-1-0325); NSF (CCF1917819); C3.ai, Inc. and Microsoft Corp.; and Google, LLC. for their generous support.

**Box 1: Engineered vs. Evolved Systems**

The most glaring difference between biological defense systems and cyber systems is how they have arisen: One system was produced by a natural evolutionary process and the other by human ingenuity. We argue that the division between these two processes is ambiguous, that modern engineering processes have more in common with evolutionary processes than is commonly believed, and that inadvertent evolutionary dynamics are particularly relevant in computer security. At first glance, the goal-directed nature of engineering, with designs produced by intelligent beings, is quite different from biological evolution, where natural selection responds to undirected random variations and drift. For example, Jacob (32) argues that evolution through natural selection is akin to tinkering and fundamentally different from the work of the master craftsman: "The engineer works according to a preconceived plan in that he foresees the product of his efforts," "The objects produced by the engineer approach the level of perfection made possible by the technology of the time." But no one would argue that today's computer systems approach perfection, nor that our software infrastructure, which is so vulnerable to attack, was produced according to a preconceived plan, even if, as humans, we can indeed foresee some futures.

In practice, engineering and evolution share many features, and it is often challenging to distinguish between the two. Many of today's engineered systems were produced at least in part by natural evolutionary processes. An obvious example is Arnold's Nobel Prize winning work using directed mutation in chemistry to optimize protein function (33). Similarly, in computing, tinkering is the norm, and clean slate design is unusual. That is, we rarely get to go back in time and redesign systems from scratch. Why? Many systems are required to maintain backward compatibility, both for communication and networking and also for user experience; it is more expensive and error-prone to redesign from scratch than to reuse existing components. This is similar to evolutionary processes, which can only "work" (evolve) with components and processes already in place, the very arguments that underlie Jacob's thesis. Despite these constraints, evolutionary processes sometimes create large shifts that can be seen on the macro scale in punctuated equilibrium (34) and on the micro scale in microbes that evolve the ability to digest new carbon sources (35)—more akin to the large-scale shifts we might associate with foresight and design, but that require neither.

We hypothesize that simple inspection of an artifact cannot always reveal the process that produced it and that at best we can make a probabilistic guess, which prompts us to ask: What are the distinct properties of engineered and evolved systems that are reflected in the designs they produce? One can even imagine a kind of Turing test that asks how one could distinguish between a product of an evolutionary process versus an engineered process? What are the hallmarks of each? Suppose, for example, that you were presented with an immune system, a cryptography system, and a modern enterprise software system with all of its defenses, would you be able to distinguish whether each was evolved or engineered?

**Box 2: Heterogeneity as Defense**

In many adversarial situations, it pays to be unpredictable. For example, an attacker that can predict defensive strategy has an advantage. Both biological and computer systems use heterogeneity as a defensive strategy to reduce predictability. In temporal heterogeneity, either the defenses or the targets themselves change dynamically, which is effective in repeated interactions against adversaries who can learn from experience. The protozoan *Trypanosoma brucei* uses this strategy to evade immune defenses, cloaking itself with one of many possible glycoproteins, and then switching the composition of its coat once its host's immune system has learned to recognize it (36). Cybersecurity has also discovered this principle and devised a variety of *moving target defenses* (37, 38, 39). Self-regulating and learning techniques also permit systems to change in response to environmental shifts, introducing additional variability.

It is not always feasible for an organism or computer system to change its structure, defenses, or even behavior during its lifetime. An alternative is to deploy heterogeneity across a population, so every individual is unique and an adversary can't predict which particular defense the target is using, thereby setting up an information asymmetry between host and pathogen. Both biology and cybersecurity have many interesting strategies for achieving this form of heterogeneity.

For example, in addition to the MHC diversity discussed in the main text, the vertebrate adaptive immune system uses somatic recombination to generate unique T-cell and B-cell receptors which are used to recognize pathogens (40). This hinders an evolving population of pathogens from anticipating which of its peptides is likely visible to immune surveillance. At the single-cell level, restriction modification systems in bacteria defend against bacteriophages (viruses that target bacteria), but they vary among bacteria. If the attacking phage DNA was synthesized in a cell with a different restriction modification system from the host (or no such system), then it is cleaved by the host's defense system and inactivated (41).

In addition to defenses to protect individuals like bacteria or vertebrates, heterogeneous defenses can be found at the species level. For example, Eurasian cuckoos produce eggs that visually mimic those of diverse songbird species, and they use this resemblance to trick songbirds into nurturing their eggs (*brood parasitism*). As a first level defense, the songbirds have evolved complex and highly divergent species-specific color patterning of their eggs. Cuckoos appear to have overcome this defense, and it appears that some songbird species have in turn evolved so each bird produces eggs with subtle individual-specific visual features that distinguish its layings from those of its conspecifics. This leaves the female cuckoo with the nearly intractable problem of mimicking the precise visual patterns of eggs in a nest it has yet to see until it is ready to lay (42).

Cybersecurity is assisted by some forms of heterogeneity that were not intentionally developed for that purpose. The diversity of hardware platforms, operating systems, and software packages may impede malicious exploitation, even though it was not deliberately engineered as a security feature. Economic and technical drivers encourage standardization and work against heterogeneity, but this unintentional diversity provides some degree of protection from widespread attacks, particularly to the users of uncommon systems (20). Directly engineered forms of heterogeneity are much more common and are effective when they create information asymmetries that favor the target against the attacker, as in cryptography systems. For example, one-time pads use a unique encryption key for every communication, which not only makes an individual message intractable to decrypt, but even if the code is broken for one message, it is not transferrable to others. Other forms of engineered heterogeneity such as N-variant systems (43) address space randomization and other binary rewriting techniques, instruction set randomization (44), platform diversity (38), all introduce heterogeneity, sometimes 'genetically' by altering each code copy or code layout, and sometimes temporally either by varying the execution conditions or changing the executing code on the fly.

In summary, the alignment between cybersecurity and biology is much closer here than for other aspects of defense. However, biology deploys diversity defenses more ubiquitously and spans more organizational layers than cybersecurity, suggesting that in the future we can expect to see new methods of heterogeneity deployed much more widely and in many more layers of the computational stack.

**Box 3: Third Party Case Studies*: Vectors in Malaria and Malware Transmission**

Does the complex biosecurity ecosystem offer insights that could be helpful in managing cyber threats?   Consider malaria, a mosquito-borne disease and one of the leading causes of morbidity and mortality in many developing countries (45).  Since the disease agent (the *plasmodium falciparum* parasite) is transmitted to humans by mosquitoes, specifically *Anopheles spp.*, the infectious disease risks to people depend on the likelihood of being bitten by infected mosquitoes. The critical factors are environmental conditions (temperature, precipitation, etc.), land use and land cover (mosquito breeding habitat), mosquito exposure to infected hosts, and defensive measures by land managers, hosts, and health authorities (46).

Local malaria control programs can act on five levels to protect a human population from malaria epidemics: 1. attack the disease agent with antimalarial drugs, 2. attack the mosquito using insecticides and reducing suitable mosquito habitats or reducing their reproductive success through genetic modifications, 3. increasing host defense, e.g., by vaccinating the population, 4. reduce exposure of mosquitoes to the pathogen via screens, bed nets, repellents, and transmission-blocking vaccines (not  developed yet), and 5. reduce contact between infected mosquitoes and people via screens, bed nets and repellents. Insecticide treated bed nets have proved particularly effective (47). They operate on multiple layers to: 1. protect uninfected people from being bitten by a malaria-infected mosquito, 2. prevent a malaria-infected person from transmitting the parasite to mosquitoes, and 3. reduce the mosquito population size (both infected and uninfected).

Local malaria risk reduction requires behavioral change on the part of land managers and potential hosts, diffusion and adoption of defensive technologies, and a set of public health and medical interventions.  Vector control operations aside, all measures depend on incentives to land managers and  hosts—the carrots and sticks offered by public health regulations and programs (48. 49).  In the same way, containment of the spread of malaria or other mosquito-borne diseases from one location to another involves restrictions on the travel of malaria-infected people, or on trade involving commodities or vessels that move mosquito eggs, larvae, or adults (49).  Where adoption of effective technologies such as treated bed nets is widespread, malaria risk has been significantly reduced (50).  Persistence of malaria elsewhere indicates that the third-party incentives alone are often insufficient to control spread infection.

There are many analogues in cybersecurity to the vectors in vector-borne diseases.  These include both infected devices, such as memory drives or CD ROMs, and infected software downloads, which transmit malicious code to individual computers or networks.  The analogue to the wider disease ecosystem includes those engaged in the manufacture and distribution of both devices and software, system administrators who maintain computer systems and are often the first to detect attacks, the originators of infected devices, and end users who invariably favor convenience and cost over security and privacy.  Most large institutions today deploy enterprise systems and compel their staff and customers to use them.  These are expensive, can be difficult to maintain or customize, and they are often tempting targets for attackers, as we have seen repeatedly with large data breaches.  If there is a lesson from malaria in such cases, it is that control is likely to be most effective when two conditions are met.  First, there needs to be an effective vector control system in place which tests for and prosecutes infected vectors.  And second, organizations and end users must adopt defensive protocols that reduce exposure to the vectors.  In cases where there are insufficient resources to implement a vector control  system, adoption of defensive protocols by end users would be critical.